\documentclass[journal=jacsat,manuscript=article]{achemso}
\usepackage{achemso}
\setkeys{acs}{maxauthors = 10}
\usepackage[version=3]{mhchem}
\usepackage[english]{babel}
\usepackage{filecontents}
\usepackage{makecell}
\usepackage{graphicx}
\usepackage{courier}
\usepackage{amssymb,amsmath,amsthm,mathrsfs,comment,afterpage,accents}
\usepackage{mathtools}
\usepackage{braket}
\usepackage{bm}
\usepackage{mathrsfs}
\usepackage{courier}
\usepackage{color}
\usepackage{subdepth}
\usepackage{tablefootnote}
\usepackage{hyperref}
\usepackage[capitalize]{cleveref}
\usepackage[linesnumbered,ruled,vlined]{algorithm2e}
\setkeys{acs}{articletitle = true}
\makeatletter
\SectionNumbersOn
\def\@dotsep{4.5}
\usepackage{cancel}
\makeatother
\usepackage{dcolumn}
\usepackage{bm}
\newcommand\mat\mathbf

\newcommand{\insertrev}[1]{{\textcolor{black} {#1}}}

\newcommand{\joonho}[1]{{\textcolor{black}{#1}}}

\title{
Stochastic Resolution-of-the-Identity Auxiliary-Field Quantum Monte Carlo: Scaling Reduction without Overhead
}

\author{Joonho Lee}
\email{linusjoonho@gmail.com}
\affiliation{
Department of Chemistry, Columbia University, 3000 Broadway, New York, New York 10027, USA
}
\author{David R. Reichman}
\affiliation{
Department of Chemistry, Columbia University, 3000 Broadway, New York, New York 10027, USA
}
\begin{document}
\newpage
\maketitle
\begin{abstract}
We explore the use of 
the stochastic resolution-of-the-identity (sRI) with
the phaseless auxiliary-field quantum Monte Carlo (ph-AFQMC) method.
sRI is combined with four existing local energy evaluation strategies in ph-AFQMC, namely (1) the half-rotated electron repulsion integral tensor (HR), (2) Cholesky decomposition (CD), (3) tensor hypercontraction (THC), or (4) low-rank factorization (LR).
We demonstrate that HR-sRI achieves no scaling reduction, CD-sRI scales as $\mathcal O(N^3)$, and THC-sRI and LR-sRI scale as $\mathcal O(N^2)$, albeit with a potentially large prefactor. 
Furthermore, the walker-specific extra memory requirement in CD is reduced from $\mathcal O(N^3)$ to $\mathcal O(N^2)$ with sRI, while sRI-based THC and LR algorithms lead to a reduction from $\mathcal O(N^2)$ extra memory to $\mathcal O(N)$. Based on numerical results for
one-dimensional hydrogen chains and water clusters,
we demonstrated that, along with the use of a variance reduction technique, CD-sRI achieves cubic-scaling {\it without overhead}.
In particular, we find for the systems studied the observed scaling of standard CD is $\mathcal O(N^{3-4})$ while for CD-sRI it is reduced to $\mathcal O(N^{2-3})$.
Once a memory bottleneck is reached, we expect THC-sRI and LR-sRI to be preferred methods due to their quadratic-scaling memory requirements and their quadratic-scaling of the local energy evaluation (with a potentially large prefactor).
The theoretical framework developed here should
facilitate large-scale ph-AFQMC applications
that were previously difficult or impossible to carry out with standard computational resources.
\end{abstract}
\newpage

\section{Introduction}
The accurate {\em ab initio} simulation of the ground state properties of molecules and solids is essential for the understanding of major swaths of chemistry and physics.  All known techniques for the reliable calculation of electronic structure are challenged by large systems containing strongly interacting electrons. Here, mean field approaches fail, and the exponential scaling of a brute force solution of the Schr\"{o}dinger equation renders the description of all but the smallest systems impossible. A large array of powerful methods have been developed which have been successfully employed in the study of large correlated systems\cite{georges1996dynamical,foulkes2001quantum,bartlett2007coupled,sun2016quantum}. Among the many available electronic structure methodologies with which to attack such systems,
quantum Monte Carlo (QMC) stands out as a unique tool
for the calculation of ground state energies
due to its attractive combination of scalability and accuracy.\cite{foulkes2001quantum}

While there are many flavors of QMC\cite{barker1979quantum,umrigar1993diffusion,ceperley1995path,foulkes2001quantum}, our focus in this work centers on the auxiliary-field QMC (AFQMC) approach.\cite{zhang1995constrained,zhang1997constrained,zhang2003quantum} AFQMC is a variant of projector QMC which is formulated
in a second-quantized determinant space.
The unbiased version of AFQMC, often referred to as free-projection AFQMC, is exact in principle, but 
has an exponential scaling with system size due to noise growth caused by the fermionic sign (or phase) problem\cite{zhang2003quantum}.
Zhang and co-workers have developed the phaseless approximation to AFQMC (ph-AFQMC), which has become
a practical and accurate tool for the {\it ab initio} simulation of molecules and materials.  
In ph-AFQMC, the phase of walker wavefunctions during imaginary-time propagation is constrained by the {\it a priori} chosen
trial wavefunction, $|\Psi_T\rangle$.  While this approach is biased, it can yield answers that systematically approach the exact solution either by the release of the constraint\cite{shi2013symmetry} or via the choice of progressively more sophisticated trial functions\cite{landinez2019non}.

In standard AFQMC simulations, walker propagation scales as $\mathcal O(OM^2)$ where $O$ is the number of electrons and $M$
is the number of single-particle basis functions.
Such cubic-scaling propagation per sample makes ph-AFQMC attractive for general {\it ab initio} problems. While there have been many successful ph-AFQMC applications to {\it ab initio} problems\cite{Al-Saidi2006,al2006auxiliary,suewattana2007phaseless,Purwanto2008,Purwanto2015,hao2018accurate,Shee2019,shee2019achieving,lee_2019_UEG,motta_kpoint,motta_forces,motta_back_prop,zhang_nio,malone_isdf,lee2020utilizing}, 
there are several remaining challenges that still need to be addressed before ph-AFQMC can become a universal tool for the study of large scale correlated electronic structure problems.  The challenge that we address in this work is the steep cost of the local energy evaluation
that is necessary for estimating the ph-AFQMC energy.
The local energy of a walker with a wavefunction $|\psi\rangle$ is given by
\begin{equation}
E_L[\psi]
=
\frac
{\langle \Psi_T | \hat{H} | \psi\rangle}
{\langle \Psi_T | \psi\rangle}\;  .
\label{def:eloc}
\end{equation}
The cubic-scaling of the walker propagation mentioned above
is asymptotically irrelevant in 
standard ph-AFQMC calculations because
the local energy evaluation scales {\it quartically} with system size without exploiting sparsity or low-rank structure.
This quartic complexity arises from the two-electron repulsion integral (ERI) tensor,
\begin{equation}
(\mu\nu|\lambda\sigma)
=
\int\mathrm{d}\mathbf r_1
\int\mathrm{d}\mathbf r_2
\frac{
\phi_\mu^*(\mathbf r_1)
\phi_\nu(\mathbf r_1)
\phi_\lambda^*(\mathbf r_2)
\phi_\sigma(\mathbf r_2)
}{|\mathbf r_1 - \mathbf r_2|}\;,
\
\end{equation}
where $\{\phi_\mu\}$ are the underlying single-particle basis functions.
In ph-AFQMC, this integral is most commonly factorized via the Cholesky decomposition (CD) into
\begin{equation}
(\mu\nu|\lambda\sigma)
 = \sum_P
L_{\mu\nu}^P 
(L_{\sigma\lambda}^P)^*\;  ,
\label{eq:cd}
\end{equation}
where $\mathbf L$ denotes a Cholesky vector. Even with CD, the evaluation of the local energy
remains a quartic-scaling task in ph-AFQMC.
\joonho{For systems with translational symmetry, it is possible to work with planewaves 
and achieve cubic-scaling with\cite{suewattana2007phaseless} or without\cite{lee_2019_UEG} fast Fourier transform.
In this work, however, we focus on general basis functions that do not exploit such symmetry.}

We mention two notable previous approaches that may be used to address this problem. The first is the tensor hypercontraction decomposition (THC)\cite{thc1,thc2,thc3} proposed by Mart{\'i}nez, Sherill, and co-workers to factorize the ERI tensor into
\begin{equation}
(\mu\nu|\lambda\sigma)
= \sum_{\hat{P}\hat{Q}}
(\eta_\mu^{\hat{P}})^*
\eta_\nu^{\hat{P}}
M_{\hat{P}\hat{Q}}
(\eta_\lambda^{\hat{Q}})^*
\eta_\sigma^{\hat{Q}}\; .
\label{eq:thc}
\end{equation}
Malone {\em et al.} have successfully applied the THC factorization of the ERI tensor to ph-AFQMC simulations and have shown that the local energy evaluation can be brought down to $\mathcal O(c_\text{THC}^2OM^2)$ 
where the value of $c_\text{THC}$ was found to be $\sim$ 8 for accurate local energy evaluation\cite{malone_isdf}. While the improved asymptotic scaling is satisfying, the resulting algorithm has a steep overhead such that the actual crossover between the conventional algorithm and the THC variant occurs for very large system sizes in practice\cite{malone_isdf,lee2019systematically}.

An alternative approach has been proposed by Motta {\em et al.}\cite{motta_thc}.  In this formulation, the nested diagonalization of Cholesky vectors proposed by Peng and Kowalski\cite{peng2017highly} is used to exploit the underlying low-rank structure of the ERI tensor. We refer this factorization as the low-rank (LR) factorization. In the LR factorization, one writes
\begin{equation}
(\mu\nu|\lambda\sigma) = \sum_{\alpha\beta}
(X_{\mu\alpha}^P)^*
U_{\nu\alpha}^P
(X_{\lambda\beta}^P)^*
U_{\sigma\beta}^P \; ,
\label{eq:lr}
\end{equation}
which arises from 
\begin{equation}
L_{\mu\nu}^P = \sum_\alpha 
(X_{\mu\alpha}^P)^*
U_{\nu\alpha}^P \; .
\label{eq:lr}
\end{equation}
This factorization achieves an asymptotically cubic-scaling local energy evaluation due to the fact that
the number of terms in the summation over $\alpha$ in \cref{eq:lr} is limited to ${\mathcal O}(\log N)$ where $N$ is the system size.
Again, due to a large overhead, this asymptotic scaling will generally not be achieved until the system reaches $\gtrsim$ 1000 electrons\cite{motta_thc}. 

The THC and LR factorization schemes enable a cubic-scaling local energy evaluation algorithm, but the overhead associated with both approaches is large enough that the actual efficiency crossover will generally not be observed for medium-sized molecules without obvious sparsity or low-rank structure. We note, however, that these algorithms offer quadratic scaling memory requirements,  which results in a reduction of the usual cubic to quartic memory requirement of standard ph-AFQMC.  In fact, such a reduction in the memory cost may be the biggest benefit gained from using the THC and LR factorizations.  We will return to a discussion of the memory reduction afforded by these techniques before concluding.

Given the above facts, it is clear that there is a need for an {\em overhead-free} cubic-scaling algorithm which can accelerate {\it ab initio} ph-AFQMC simulations for medium-sized systems.  Based on earlier work of Baer and Neuhauser\cite{baer2012communication},  
Neuhauser, Baer, Rabani and co-workers have developed the use of stochastic orbitals as a means of reducing scaling in a variety of electronic structure methodologies, including DFT\cite{baer2013self,neuhauser2016stochastic,neuhauser2014communication}, TDDFT\cite{gao2015sublinear}, second-order M{\o}ller-Plesset perturbation theory (MP2) \cite{neuhauser2013expeditious,ge2014guided,takeshita2017stochastic}, second-order perturbative Green's functions (GF2)\cite{neuhauser2017stochastic,takeshita2019stochastic,dou2019stochastic}, the random phase approximation\cite{neuhauser2013expeditiousrpa}, GW\cite{rabani2015time}, and the Bethe-Salpeter equation\cite{rabani2015time}.

Drawing from this body of work, we explore the combined use of the 
stochastic resolution-of-the-identity (sRI) formulation of Takeshita {\em et al.}\cite{takeshita2017stochastic} with ph-AFQMC. 
\insertrev{The central idea underlying sRI is a stochastic resolution of the Coulomb operator which was first developed in ref. \citenum{neuhauser2016stochastic}.}
We will demonstrate that sRI can be naturally incorporated into ph-AFQMC and used to reduce the
formal scaling of the local energy evaluation
to cubic-scaling when combined with CD, and to quadratic-scaling when combined with THC or LR factorization.
Most notably, we will numerically demonstrate that the combined use of Cholesky decomposition and the stochastic resolution of the identity (CD-sRI) with a simple variance reduction technique allows for a cubic-scaling algorithm {\it without overhead}. Given this attractive feature, we expect the CD-sRI approach to become the standard method for the computation of such quantities for medium-sized systems, and THC-sRI and LC-sRI formulations to be the preferred approach for very large ones.

This paper is organized as follows:
(1) we give a brief review of the ph-AFQMC algorithm, 
(2) we analyze formal scaling of existing local energy evaluation strategies, 
(3) we examine scaling reduction in these strategies when combined with sRI,
and
(4) we present numerical examples (hydrogen chains and water clusters) to show
speed-up and scaling reduction of the CD approach with sRI.

\section{Theory} 
To set precise notation for our discussion of scaling, in the following we use $O$ to denote the number of occupied orbitals, 
$M$ to denote the number of single-particle basis functions,
$X$ to denote the number of Cholesky vectors (or auxiliary basis functions),
and $N$ to denote the system size in general.
\subsection{Review of Auxiliary-Field Quantum Monte Carlo}
We briefly review the basic algorithmic details of AFQMC here and refer interested readers to ref. \citenum{Motta2019} for a more extended review of modern AFQMC development.
We start from an {\it ab initio} Hamiltonian $\hat H$ given by
\begin{equation}
\hat{H} = 
\sum_{pq} h_{pq}\hat{a}_p^\dagger \hat{a}_q
+
\frac12
\sum_{pqrs}  
(pr|qs)
\hat{a}_p^\dagger 
\hat{a}_q^\dagger 
\hat{a}_s
\hat{a}_r
\equiv \hat{H}_1 + \hat{H}_2\; ,
\end{equation}
where $\hat{H}_1$ and $\hat{H}_2$ are the one-body and two-body part of $\hat{H}$, respectively.
As with other projector QMC methods, AFQMC starts from
\begin{equation}
|\Psi_0\rangle 
\propto
\lim_{\tau \rightarrow\infty}e^{-\tau\hat{H}} | \Phi_0\rangle \; ,
\label{eq:pmc}
\end{equation}
where $\hat{H}$ is the Hamiltonian, $\tau$ denotes imaginary time, $|\Psi_0\rangle$ is the exact ground state, and $|\Phi_0\rangle$ can be any wavefunction with a non-zero overlap with the true ground state.
A key feature of AFQMC is that one employs the Hubbard-Stratonovich transformation along with the Trotter approximation to simplify the many-body propagator in \cref{eq:pmc} and reduce the problem to that of a series of one-body problems in a fluctuating auxiliary field.
With the Trotter approximation 
for a given time step $\Delta \tau$,
the propagator is expressed as
\begin{equation}
e^{-\Delta\tau\hat{H}}
=
e^{-\Delta\frac{\tau}{2}\hat{H}_1}
e^{-\Delta\tau\hat{H}_2}
e^{-\Delta\frac{\tau}{2}\hat{H}_1}
+ \mathcal O (\Delta \tau ^2)\; .
\end{equation}
The Hubbard-Stratonovich transformation allows for the rewriting of the two-body propagator as
an integration over auxiliary fields $\mathbf x$,
\begin{equation}
e^{-\Delta\tau\hat{H}_2}
=
\int \mathrm{d}{\mathbf x}
\:
p(\mathbf{x})
e^{-\sqrt{\Delta\tau} \mathbf{x}\cdot\hat{\mathbf{v}}}\; ,
\end{equation}
where the one-body operator $\hat{\mathbf{v}}$ is obtained from
\begin{equation}
    \hat{H}_2 = -\frac12 \sum_P^{X} \hat{v}_P^2 \; .
\label{eq:sq}
\end{equation}
\cref{eq:sq} is usually achieved via the CD of the ERI tensor. Alternatively, density-fitting can be used.
With this decomposition, the total propagator for given fields $\mathbf x$ and a given time step $\Delta \tau$ is written as
\begin{equation}
\hat{B}(\Delta \tau, \mathbf x) = e^{-\frac{\Delta\tau}{2} \hat{H}_1}
e^{-\sqrt{\Delta\tau} \mathbf{x}\cdot\hat{\mathbf{v}}}
e^{-\frac{\Delta\tau}{2} \hat{H}_1} \; .
\label{eq:B}
\end{equation}
By the virtue of the Thouless theorem\cite{thouless1960stability,thouless1961vibrational}, 
the application of this propagator to a single determinant
remains a single determinant. This allows for an efficient walker propagation in AFQMC.

\insertrev{One can reduce statistical fluctuations greatly by employing a mean-field subtraction\cite{Motta2019} which is closely related to the shifted contour technique developed by Neuhauser and co-workers\cite{rom1997shifted}.
The mean-field subtraction technique redefines the one-body and two-body using the expectation value of $\hat{v}_P$ with a trial wavefunction,
\begin{equation}
\langle\hat{{v}}_P\rangle_T \equiv \langle \Psi_T | \hat{ v}_P | \Psi_T \rangle
\end{equation}
We then define
\begin{align}
\hat{H}_1' &= \hat{H}_1 - \sum_P^X \hat{v}_P \langle\hat{{v}}_P\rangle_T + \frac12 \sum_P^{X} \langle\hat{{v}}_P\rangle_T^2  \\
    \hat{H}_2' &= -\frac12 \sum_P^{X} (\hat{v}_P - \langle\hat{{v}}_P\rangle_T) ^2 ,
\end{align}
which maintains $\hat{H} = \hat{H}_1' + \hat{H}_2'$.
With this, the propagator in \cref{eq:B} uses $\hat{H}_1'$ and $\hat{\mathbf v} - \langle\hat{\mathbf{v}}\rangle_T$ instead of $\hat{H}_1$ and $\hat{\mathbf v}$. 
This simple subtraction has been shown to be effective in reducing statistical fluctuations \cite{rom1997shifted, Motta2019} and this is what is used throughout this work.
}

The imaginary-time equation-of-motion is dictated by the propagator in \cref{eq:B}.
With importance sampling via a trial wavefunction $|\Psi_T\rangle$, 
we write the global AFQMC wavefunction at imaginary-time $\tau$ as
\begin{equation}
|\Psi(\tau)\rangle
=
\sum_i^{N_w}
w_i (\tau)
\frac{|\psi_i(\tau)\rangle}
{\langle \Psi_T | \psi_i(\tau)\rangle}\; ,
\end{equation}
where $N_w$ is the number of walkers.
Within the phaseless approximation\cite{zhang2003quantum},
we update the $i$-th walker weight and determinant via
\begin{align}
|\psi_i(\tau+\Delta\tau)\rangle &= \hat{B}(\Delta\tau, \mathbf{x}_i-\mathbf{\bar{x}}_i) |\psi_i(\tau)\rangle \; ,\\
w_i(\tau+\Delta\tau) &= I_{ph}(\mathbf{x}_i,\mathbf{\bar{x}}_i,\tau,\Delta\tau) \times w_i(\tau)\; ,
\end{align}
where the force-bias, $\mathbf{\bar{x}}_i$, is defined as
\begin{equation}
    \mathbf{\bar{x}}_i(\Delta \tau, \tau) = -\sqrt{\Delta\tau}
\frac{\langle
    \Psi_T | \hat{\mathbf{v}} - \langle\hat{\mathbf{v}}\rangle_T | \psi_i(\tau)
    \rangle}{
    \langle
    \Psi_T | \psi_i(\tau)
    \rangle
    }\; .
\end{equation}
The phaseless importance function is defined as
 \begin{equation}
I_{ph}(\mathbf{x}_i,\mathbf{\bar{x}}_i,\tau,\Delta\tau) = |I (\mathbf{x}_i,\mathbf{\bar{x}}_i,\tau,\Delta\tau)|\times
\text{max}(0, \cos(\theta_i(\tau)))\; ,
\label{eq:ph}
\end{equation}
with the hybrid importance function given by
\begin{equation}
    I(\mathbf{x}_i,\mathbf{\bar{x}}_i,\tau,\Delta\tau) = S_i(\tau, \Delta\tau)
    e^{\mathbf{x}_i\cdot\mathbf{\bar{x}}_i-\mathbf{\bar{x}}_i\cdot\mathbf{\bar{x}}_i/2}\; ,
 \label{eq:import}
\end{equation}
the overlap ratio of the $i$-th walker $S_i$ give by
\begin{equation}
S_i(\tau, \Delta\tau) = \frac{\langle
    \Psi_T |
    \hat{B}(\Delta\tau, \mathbf{x}_i-\mathbf{\bar{x}}_i) | \psi_i(\tau)
    \rangle}{
    \langle
    \Psi_T | \psi_i(\tau)
    \rangle}\; ,
 \label{eq:ovl}
\end{equation}
and the phase $\theta_i(\tau)$ is given by
\begin{equation}
\theta_i(\tau) = \text{arg}\left(
S_i(\tau, \Delta\tau)
\right)\; .
\label{eq:theta}
\end{equation}
This completes our compact description of the ph-AFQMC algorithm.

\subsection{Existing Strategies for the Local Energy Evaluation}
In practical AFQMC calculations one usually targets the ground state energy estimated via 
\begin{equation}
\langle E \rangle = 
\frac{\sum_i w_i E_L[\psi_i]}{\sum_i w_i}\; ,
\label{eq:elocal}
\end{equation}
where $i$ indexes the $i$-th walker, $w_i$ is the weight of the $i$-th walker, and $E_L[\psi_i]$ is the local energy of the $i$-th walker
as defined in \cref{def:eloc}.
Without any approximations, the two-body contribution to the local energy is, via Wick's theorem, given by 
\begin{equation}
E_L^{[2]} [\psi]= \frac12 \sum_{pqrs} (pr|qs) (G_{pr}G_{qs} - G_{ps}G_{qr})\; ,
\label{eq:elocalref}
\end{equation}
where the Green's function $\mathbf G$ is a walker-dependent quantity defined as
\begin{equation}
G_{pr} = 
\frac
{\langle \Psi_T | \hat{a}_p^\dagger \hat{a}_r | \psi \rangle}
{\langle \Psi_T | \psi \rangle}
= (\Theta (\mathbf C_T)^\dagger)_{rp}\; .
\end{equation}
Here, $\mathbf C_T$ is the molecular orbital (MO) coefficient matrix for the trial wavefunction and
$\Theta$ is defined as
\begin{equation}
\Theta = \mathbf C (\mathbf C_T^\dagger \mathbf C)^{-1} \; ,
\end{equation}
with $\mathbf C$ the MO coefficient matrix for the walker determinant $|\psi\rangle$.

Without exploiting any structure in the ERI tensor, the local energy evaluation scales quartically with system size.
For instance, a standard way to evaluate \cref{def:eloc} is to use so-called a ``half-rotated'' (HR) ERI tensor \cite{al2006auxiliary,malone_isdf},
\begin{equation}
E_{L,\text{HR}}^{[2]} [\psi]= \frac12 \sum_{ijrs} (ir|js) (\Theta_{ri}\Theta_{sj} - \Theta_{si}\Theta_{rj})
\label{eq:elocal1}
\end{equation}
while storing $(ir|js)$ in memory. This algorithm requires $\mathcal O(O^2 M^2)$ memory and scales as $\mathcal O(O^2 M^2)$.
\joonho{Each walker additionally requires $\mathcal O(O^2 M)$ memory to save intermediates that appear when evaluating \cref{eq:elocal1}.}
Furthermore, the formation of $(ir|js)$ at the beginning of the QMC run scales as $\mathcal O(O M^4)$ which is more expensive than the local energy evaluation, although it needs to be performed only once.
Despite these relatively steep scaling behaviors, HR is perhaps the most widely used local energy algorithm when the memory cost is affordable. 
We refer this algorithm to as the ``half-rotated'' (HR) local energy evaluation.

Another standard evaluation procedure is to work with density-fitted integrals or Cholesky vectors directly.
This may be achieved by using \cref{eq:cd} within the local energy expression, yielding
\begin{equation}
E_{L,\text{CD}}^{[2]} [\psi]= \frac12 \sum_{ijrs} \sum_P (L_{ir}^P(L_{sj}^P)^*) (\Theta_{ri}\Theta_{sj} - \Theta_{si}\Theta_{rj})\; ,
\label{eq:elocal2}
\end{equation}
which 
scales as $\mathcal O(O^2 M X)$.
In general $X$ is 4-5 times larger than $M$ and therefore \cref{eq:elocal2} is slower than \cref{eq:elocal1}.
However, \cref{eq:elocal2} requires only $\mathcal O(OMX)$ memory storage which makes it better suited for large-scale simulations.
\joonho{To save intermediates, each walker additionally requires $\mathcal O(O^2 X)$ memory.}
We refer this algorithm to as the ``Cholesky decomposition'' (CD) local energy evaluation.

Lastly, we note that some recent advances in integral factorization techniques have allowed for a cubic-scaling local energy evaluation with a quadratic memory storage\cite{malone_isdf}.
We first discuss the THC strategy\cite{thc1,thc2,thc3} invented by Sherill, Mart{\'i}nez and coworkers, which was applied to AFQMC by Malone {\em et al.}\cite{malone_isdf}.
Specifically, we use \cref{eq:thc} within the local energy evaluation, yielding
\begin{equation}
E_{L,\text{THC}}^{[2]} [\psi]= \frac12 \sum_{ijrs}
\left(
\sum_{\hat{P}\hat{Q}}
(\eta_i^{\hat{P}})^*
\eta_r^{\hat{P}}
M_{\hat{P}\hat{Q}}
(\eta_j^{\hat{Q}})^*
\eta_s^{\hat{Q}}
\right)
(\Theta_{ri}\Theta_{sj} - \Theta_{si}\Theta_{rj})\; .
\label{eq:elocal3}
\end{equation}
All of the tensor contractions in \cref{eq:elocal3} are simple matrix-matrix multiplications, leading to cubic scaling with system size. More precisely, the local energy calculation scales as $\mathcal O(c_\text{THC}^2OM^2)$, where we define the THC rank to be $c_\text{THC}M$.
The memory requirement is set by storing $\eta_p^{\hat{P}}$ and $\mathbf M$ which have only a quadratic number of non-zero values. 
\joonho{Furthermore, each walker additionally requires $\mathcal O(c_\text{THC}^2M^2)$ memory to save intermediates that naturally appear in \cref{eq:elocal3}.}
While these asymptotic properties are highly attractive, the THC algorithm has been shown to have a large prefactor compared to \cref{eq:elocal1} and \cref{eq:elocal2} because $c_\text{THC}$ can be quite large in practice\cite{malone_isdf}. For example, $c_\text{THC}$ was found to be approximately 8 for the accurate calculation of the local energy of diamond within the double-zeta basis.

Similarly, for the LR factorization of Motta {\em et al.} \cite{motta_thc}, we use \cref{eq:lr} with the AFQMC local energy expression
yielding
\begin{equation}
E_{L,\text{LR}}^{[2]} [\psi]= \frac12 \sum_{ijrs}
\left(
\sum_{\alpha\beta}
(X_{i\alpha}^P)^*
U_{r\alpha}^P
(X_{j\beta}^P)^*
U_{s\beta}^P
\right)
(\Theta_{ri}\Theta_{sj} - \Theta_{si}\Theta_{rj})\; ,
\label{eq:elocal4}
\end{equation}
which scales as
$\mathcal O(n_r OMX)$ where the rank $n_r$ sets the upper limit on the summation over $\alpha$ and $\beta$ and scales like $\log (N)$. Note that this approach also effectively achieves a cubic-scaling local energy evaluation algorithm.
\joonho{The memory requirement is $\mathcal O(n_r O X + n_r M X)$ and each walker additionally requires $\mathcal O(n_r OX)$ to save intermediates.}
However, as with the use of THC, $n_r$ is usually large and therefore for practical applications,
useful cubic-scaling may not be observed\cite{motta_thc} \joonho{and only memory saving may be practically useful}.

\subsection{Stochastic Resolution-of-the-Identity} \label{sec:localenergy}
We follow the explication of the stochastic resolution-of-the-identity (sRI) technique proposed by Takeshita {\em et al.}\cite{takeshita2017stochastic} which has been shown to lower the scaling of MP2 (RI-MP2)\cite{takeshita2017stochastic} and the second order Green's function method (RI-GF2)\cite{takeshita2019stochastic,dou2019stochastic}. 
sRI is based on the simple mathematical observation that one may represent the Kronecker delta function with stochastic functions as,
\begin{equation}
\delta_{\alpha\beta} = 
\lim_{N_\xi \rightarrow \infty}
\frac{1}{N_\xi}
\sum_{\xi=1}^{N_\xi}
\theta_{\alpha}^{\xi}
\theta_{\beta}^{\xi}\; ,
\label{eq:sri}
\end{equation}
where
$N_\xi$ is the number of stochastic samples
and
$\pmb{\theta}^{\xi}$ is an sRI basis function
whose entry is randomly chosen to be $\pm 1$.
In practice, we will limit the number of samples to a finite number which does not scale with system size. 
This restriction has been shown to be sufficient to achieve a fixed statistical error per particle in MP2 and GF2.
The key feature of this approach is that
it does not assume any structure (either sparsity or low-rank) in the underlying ERI tensor
while still reducing the cost.
Due to this fact, the overhead of this approach is almost negligible,
especially when looking at size-intensive quantities for which ph-AFQMC applications are also well-suited.

We emphasize that the scaling of a QMC algorithm
should not be discussed without assessment of the
the underlying statistical error.
For instance, one may argue that propagation within ph-AFQMC scales cubically, but
the number of samples associated with a fixed statistical error scales linearly with system size.
This then makes the ph-AFQMC propagation scale quartically with system size for a fixed statistical error.
The usual cubic scaling of the ph-AFQMC propagation quoted in literature\cite{lee_2019_UEG}
is the scaling for a fixed statistical error {\it per particle}.
Similarly, the lower scaling of sRI-MP2 and sRI-GF2 approaches cited above is only justified 
when considering observables for a fixed statistical error {\it per particle}. 
This makes the combination of these two methods very natural.
We also note that it is still possible to obtain total energies for a fixed statistical error in both the ph-AFQMC and sRI methods, albeit with an increased cost.

\subsubsection{The Half-rotated sRI (HR-sRI) algorithm}
We first re-write \cref{eq:elocal1} with two Kronecker deltas,
\begin{equation}
E_{L,\text{HR}}^{[2]} [\psi]= 
\frac12 \sum_{ijrs} 
\sum_{pq}
(ir|js)\delta_{rp}\delta_{sq} (\Theta_{pi}\Theta_{qj} - \Theta_{qi}\Theta_{pj})\; .
\end{equation}
Next, one may employ two sets of stochastic orbitals to represent these Kronecker deltas and write the local energy as
\begin{equation}
E_{L,\text{HR-sRI}}^{[2]}[\psi] \; ,
=
\frac{1}{2N_\xi^2} \sum_{ij}
\sum_{\xi\xi'}
\left[
(i\xi|j\xi')
\left(
\Theta_{\xi i} \Theta_{\xi' i}
-\Theta_{\xi' i} \Theta_{\xi i}
\right)
\right]\; ,
\end{equation}
where
\begin{equation}
(i\xi|j\xi')
= \sum_{rs}
\theta_r^\xi
\theta_s^{\xi'}
(i r | j s)\; ,
\label{eq:int1}
\end{equation}
and
\begin{equation}
\Theta_{\xi i} = 
\sum_r
\theta_r^\xi
\Theta_{r i}\; .
\end{equation}
The formation of \cref{eq:int1} is the bottleneck, scaling as
$\mathcal O (N_\xi O^2M^2)$.
\joonho{The additional memory usage of each walker scales as $\mathcal O(N_\xi O^2 M)$.}
Thus, there is no reason to employ this algorithm as the asymptotic scaling is not improved by sRI \joonho{and no memory saving is obtained}.

\subsubsection{The Cholesky decomposition sRI (CD-sRI) algorithm}\label{sec:CD-sRI}
We next insert \cref{eq:sri} into \cref{eq:elocal2} and show that the scaling of the local energy evaluation is reduced from quartic to cubic. 
The combination of the sRI expression to the CD-based local energy algorithm is most natural in the auxiliary basis function space.
Namely,
\begin{equation}
\sum_P (L_{ir}^P(L_{sj}^P)^*)
=
\sum_{PQ} (L_{ir}^P\delta_{PQ}(L_{sj}^Q)^*)
=
\frac1{N_\xi}\sum_\xi 
( \sum_P\theta_{P}^{\xi} L_{ir}^P)
( \sum_Q\theta_{Q}^{\xi} (L_{sj}^Q)^*)\; .
\end{equation}
We further define an intermediate tensor $\mathbf R$,
\begin{equation}
R_{ir}^\xi \equiv   \sum_P\theta_{P}^{\xi} L_{ir}^P \; .
\end{equation}
The formation of $\mathbf R$ scales as $\mathcal O(N_\xi OMX)$ which is cubic-scaling for a size-intensive quantity $N_\xi$.
The Coulomb term in \cref{eq:elocal2} is thus expressed as
\begin{equation}
E_{J,\text{CD}} [\psi]= \frac12 \sum_{ijrs} \sum_P (L_{ir}^P(L_{sj}^P)^*) \Theta_{ri}\Theta_{sj} \; ,
\label{eq:ecoul}
\end{equation}
which scales as $\mathcal O(OMX)$. 
\joonho{The extra memory cost for storing walker-specific intermediates
scales as
$\mathcal O(X)$.}
Using $\mathbf R$, 
\begin{equation}
E_{J,\text{CD-sRI}} [\psi]= \frac1{2N_\xi} \sum_{ijrs} \sum_\xi (R_{ir}^\xi(R_{sj}^\xi)^*) \Theta_{ri}\Theta_{sj} 
= \frac{1}{N_\xi}\sum_\xi E_{J,\text{CD-sRI}}^\xi [\psi]\; ,
\label{eq:ecoul2}
\end{equation}
where
\begin{equation}
 E_{J,\text{CD-sRI}}^\xi [\psi] \equiv
 \frac12 \sum_{ijrs} (R_{ir}^\xi(R_{sj}^\xi)^*) \Theta_{ri}\Theta_{sj} \; .
\end{equation}
The summation over $P$ is term now replaced by the summation over $\xi$, which lowers the scaling from $\mathcal O(OMX)$ to $\mathcal O(OMN_\xi)$. \joonho{The walker-specific extra memory cost is reduced from $\mathcal O(X)$ to $\mathcal O(N_\xi)$.}
The exchange term in \cref{eq:elocal2} is
\begin{equation}
E_{K,\text{CD}} [\psi]= -\frac12 \sum_{ijrs} \sum_P (L_{ir}^P(L_{sj}^P)^*) \Theta_{si}\Theta_{rj}
\label{eq:eexch}
\end{equation}
and scales as $\mathcal O(O^2MX)$ \joonho{with $\mathcal O(O^2 X)$ walker-specific extra memory cost}. This term now becomes
\begin{equation}
E_{K,\text{CD-sRI}} [\psi]= -\frac1{2N_\xi} \sum_{ijrs} \sum_\xi (R_{ir}^\xi(R_{sj}^\xi)^*) \Theta_{si}\Theta_{rj}
=\frac1{N_\xi}\sum_\xi E_{K,\text{CD-sRI}} ^\xi[\psi]\; ,
\label{eq:eexch2}
\end{equation}
where
\begin{equation}
E_{K,\text{CD-sRI}} ^\xi[\psi] \equiv -\frac12 \sum_{ijrs} (R_{ir}^\xi(R_{sj}^\xi)^*) \Theta_{si}\Theta_{rj}\; ,
\end{equation}
which scales as $\mathcal O(O^2MN_\xi)$. \joonho{Its walker-specific memory cost scales as $\mathcal O(N_\xi O^2)$.}
This completes the demonstration of a scaling reduction of \cref{eq:elocal2} to cubic scaling.
In summary, the asymptotic scaling of CD-sRI algorithm is
$\mathcal O(N_\xi OMX + N_\xi O^2 M)$ which includes the cost of the formation of $\mathbf R$ as well.
\joonho{The additional memory cost due to storing $\mathbf R$ and other intermediates scales as $\mathcal O(N_\xi OM+ N_\xi O^2)$. This is also an improvement over the conventional CD algorithm.}

\subsubsection{Tensor hypercontraction sRI (THC-sRI) algorithm}
We next apply sRI to \cref{eq:elocal3} and show that the overall scaling can be reduced to {\it quadratic}.
With the use of two sets of sRI insertions, we write
\begin{equation}
E_{L,\text{THC-sRI}}^{[2]} [\psi]= 
\frac{1}{N_\xi^2}
\sum_{\xi\xi'}
E_{L,\text{THC-sRI}}^{\xi\xi'} [\psi]\; ,
\end{equation}
with
\begin{equation}
E_{L,\text{THC-sRI}}^{\xi\xi'} [\psi]= 
\frac{1}{2} \sum_{ij}
\sum_{\xi\xi'}
\left(
\sum_{\hat{P}\hat{Q}}
(\eta_i^{\hat{P}})^*
\eta_\xi^{\hat{P}}
M_{\hat{P}\hat{Q}}
(\eta_j^{\hat{Q}})^*
\eta_{\xi'}^{\hat{Q}}
\right)
(\Theta_{\xi i}\Theta_{\xi' j} - \Theta_{\xi' i}\Theta_{\xi j})\; ,
\end{equation}
where we define
\begin{equation}
\eta_\xi^{\hat{P}} = \sum_r \eta_r^{\hat{P}} \theta_{r}^\xi \;.
\end{equation}
With a proper set of contractions, one can show that this expression scales as
$\mathcal O(N_\xi c_\text{THC} M^2+N_\xi OM + N_\xi c_\text{THC}^2 M^2 )$.
\joonho{The additional walker memory usage due to storing intermediates comes at a cost of
$\mathcal O(N_\xi c_\text{THC} M + N_\xi^2 c_\text{THC} M)$. THC-sRI improves both scaling and memory usage over the conventional THC algorithm.}
\subsubsection{Low-rank factorization sRI (LR-sRI) algorithm}
The same conclusion can be deduced for the LR-sRI algorithm by employing 
two sets of sRI expressions. With sRI, one may write \cref{eq:elocal4} as
\begin{equation}
E_{L,\text{LR-sRI}}^{[2]} [\psi]= 
\frac{1}{N_\xi^2}
\sum_{\xi\xi'}
E_{L,\text{LR-sRI}}^{\xi\xi'} [\psi]\; ,
\end{equation}
with
\begin{equation}
E_{L,\text{LR-sRI}}^{\xi\xi'} [\psi]= 
\frac{1}{2} \sum_{ij}
\left(
\sum_{\alpha\beta}
(X_{i\alpha}^P)^*
U_{\xi\alpha}^P
(X_{j\beta}^P)^*
U_{\xi'\beta}^P
\right)
(\Theta_{\xi i}\Theta_{\xi' j} - \Theta_{\xi' i}\Theta_{\xi j})\; .
\end{equation}
Here we have defined
\begin{equation}
U_{\xi\alpha}^P = \sum_{r} U_{r\alpha}^P \theta_{r}^\xi\; .
\end{equation}
With an appropriate series of contractions, 
it can be shown that this local energy evaluation scales as
$
\mathcal O(N_\xi n_r MX
+
 N_\xi OM
+
 N_\xi n_r OX)$.
 \joonho{The extra memory required for storing walker-specific intermediates
 is $\mathcal O(N_\xi n_r X)$. As with THC-sRI, we observe an improvement using LR-sRI in both scaling and memory usage over the LR algorithm.}

\subsection{Sampling of the Stochastic Resolution-of-the-Identity and the Global Energy} \label{sec:cv}
The implementation of the sRI local energy estimator 
can be achieved with only minor modifications to existing AFQMC programs. 
There are two stochastic samplings to complete the evaluation of the sRI global energy: 
one is the standard AFQMC walker local energy sampling for each walker in \cref{eq:elocal}, 
and the other 
is the sRI stochastic orbital sampling in \cref{eq:sri}.
A simple algorithm for this is as follows:
\begin{enumerate}
\item Each walker samples a set (or two sets depending on the choice of algorithms) of $N_\xi$ sRI orbitals $\pmb \theta$.
\item Evaluate the corresponding local energy expression derived in \cref{sec:localenergy}. 
\item The global mixed estimator energy is then estimated in the standard manner as in \cref{eq:elocal} as before.
\end{enumerate}

Minimizing $N_\xi$ is critical in order to achieve the aforementioned lower scaling with as little overhead as possible.
Fortunately, within the QMC set up, it is natural to set $N_\xi = 1$ per sample because
the representation power of sampled sRI orbitals increases
as stochastic samples are accumulated
throughout the imaginary-time propagation. In other words, at each time step each walker samples one sRI orbital and
the global estimator is obtained by stochastically averaging over many such samples.
In the limit of infinite statistical sampling, this algorithm converges to the exact ph-AFQMC energy (i.e. the one without sRI).

The use of 
a variance reduction (VR) technique\cite{spink2013quantum,lemieux2014control} for sRI-ph-AFQMC  has been found to be very effective.  For any of the algorithms outlined in this work, we write
\begin{equation}
E_L^{[2]}[\psi]
=
E_L^{[2]}[\psi_T] 
+
\sum_{\xi\xi'} (E_L^{\xi\xi'}[\psi]-E_L^{\xi\xi'}[\psi_T])\; ,
\label{eq:var}
\end{equation}
where $\psi_T$ is the trial wavefunction. 
As suggested by this equation, we compute the ``correlation'' contribution to the local energy via sRI and
the mean-field energy is reused.
By expressing the energy in this form, the statistical fluctuations are found to be greatly reduced.
This technique is often referred to as the ``control variate'' approach in Monte Carlo\cite{spink2013quantum,lemieux2014control}.
\insertrev{We note that the efficacy of the control variate approach in this work will ultimately depend on the quality of the trial wavefunction as well.
As long as the trial wavefunction is of good quality, the variance reduction will be effective.
}

\subsection{Summary of the sRI Algorithms}

\begin{table}[h]
\begin{tabular}{|c|c|c|c|c|}\hline
 & HR & CD   &  THC & LR \\ \hline
 conventional & $\mathcal O(O^2M^2)$ & $\mathcal O(O^2MX)$ & $\mathcal O(c_\text{THC}^2 O M^2)$ & $\mathcal O(n_r OMX)$\\ \hline
 sRI & 
 $\mathcal O(N_\xi O^2 M^2)$ & 
 $\substack{\mathcal O(N_\xi OMX \\+ N_\xi O^2 M)}$ & 
 $ \substack{\mathcal O(N_\xi c_\text{THC} M^2+N_\xi OM \\+ N_\xi c_\text{THC}^2 M^2)} $
 & 
 $
 \substack{\mathcal O(N_\xi n_r MX
+
 N_\xi OM
\\+
 N_\xi n_r OX)}
 $\\ \hline
 leading speedup & 
none & 
 $O/N_\xi$ & 
 $O/N_\xi$ & 
 $O/N_\xi$ 
 \\\hline
\end{tabular}
\caption {Computational scaling of
different algorithms for evaluating the two-body contribution to the energy as in \cref{eq:elocalref}.
Used acronyms are
HR = half-rotated scheme, CD = Cholesky decomposition scheme, THC = tensor hypercontraction approach, and LR = low-rank factorization approach.
This scaling does not include computations that occur only once at the beginning of a QMC run.
}
\label{tab:scaling}
\end{table}

We have discussed a total of four local energy evaluation strategies and
how the sRI can reduce the scaling of three of them.
We summarize
this in \cref{tab:scaling}.
With the sRI, we have formulated one cubic-scaling (CD-sRI) and two quadratic-scaling (THC-sRI and LR-sRI) algorithms.

\joonho{
\begin{table}[h]
\begin{tabular}{|c|c|c|c|c|}\hline
 & HR & CD   &  THC & LR \\ \hline
 conventional & 
 $\mathcal O(O^2M)$ & 
 $\mathcal O(O^2X)$ & 
 $\mathcal O(c_\text{THC}^2 M^2)$ & 
 $\mathcal O(n_r OX + n_r MX)$\\ \hline
 sRI & 
 $\mathcal O(N_\xi O^2 M)$ & 
 ${\mathcal O(N_\xi OM + N_\xi O^2)}$ & 
 $ \substack{\mathcal O(N_\xi c_\text{THC} M\\+N_\xi^2 c_\text{THC} M)} $
 & 
 $
{\mathcal O(
 N_\xi n_r X)}$\\ \hline
 leading saving & 
none & 
 $OX/(N_\xi M)$ & 
 $c_\text{THC} M/N_\xi^2$ & 
 $M/N_\xi$ 
 \\\hline
\end{tabular}
\caption {Additional walker-specific memory requirement of
different algorithms for evaluating the two-body contribution to the energy as in \cref{eq:elocalref}.
Used acronyms are
HR = half-rotated scheme, CD = Cholesky decomposition scheme, THC = tensor hypercontraction approach, and LR = low-rank factorization approach.
This does not include memory usage for storing shared tensors across all of the walkers.
}
\label{tab:memory}
\end{table}
For each algorithm, we have also discussed the additional memory requirement of each walker for storing intermediates.
This is summarized in \cref{tab:memory}. With sRI, we have one algorithm (CD-sRI) that requires additional quadratic-scaling memory and two algorithms that require additional linear-scaling (THC-sRI and LR-sRI) memory. These are all improvements over their conventional counterparts.
}

Among those algorithms,
we choose CD-sRI to demonstrate 
the behavior of sRI-ph-AFQMC methods
due to its simple implementation, reasonable leading speedup of $O/N_\xi$, \joonho{and leading walker-specific memory saving of $\mathcal O(OX/(N_\xi M))$}.
\insertrev{The implementation of THC-sRI and LR-sRI is straightforward if THC and LR are already implemented. However, in \texttt{PAUXY}\cite{pauxy}, these are currently unavailable, which led us to choosing CD-sRI for the purpose of demonstration.} Moreover, we apply the sRI only to the exchange contribution (i.e., $E_K$)
because the Coulomb contribution (i.e., $E_J$) can be evaluated at cubic cost within the standard algorithm.
More specifically, we write (with the VR technique)
\begin{equation}
E_{L,\text{CD-sRI}}^{[2]}[\psi]
=
E_{J,\text{CD}}[\psi]
+
E_{K,\text{CD}}[\psi_T]
+
(E_{K,\text{CD-sRI}}[\psi] - E_{K,\text{CD-sRI}}[\psi_T])\; ,
\label{eq:cdsrifinal}
\end{equation}
where $E_{K,\text{CD}}[\psi_T]$ is evaluated only once in the beginning and used throughout the simulation.
Since $E_{K,\text{CD-sRI}}$ is evaluated twice (once for each of $\psi$ and $\psi_T$), 
this adds an additional prefactor of 2. This prefactor increase is negligible compared to the variance reduction that we gain, as we shall see.
The resulting algorithm is overall cubic-scaling {\it per sample}.

\section{Computational Details}
All calculations were performed with an open-source python-based AFQMC program called \texttt{PAUXY}\cite{pauxy}.
The pair-branching algorithm was employed for the population control\cite{wagner_qwalk}. 
The time step of 0.005 a.u. was used in all computations.
We set $N_\xi = 1$ throughout all sRI calculations. \insertrev{Increasing $N_\xi$ provided almost no improvement in statistical efficiency for systems considered in this work.}
This enables a speedup of the order of $O$ (the number of electrons) compared to the conventional CD algorithm.
We use the cc-pVDZ\cite{Dunning1989} basis set in all examples presented below.
We used restricted Hartree-Fock as a trial wavefunction throughout \insertrev{because systems considered in this work are dominated mainly by dynamical correlation}. 
The truncation threshold for the Cholesky decomposition was set to $10^{-5}$.
A total of 160 walkers were used in every calculation presented below.
QMCPACK\cite{qmcpack,qmcpack2} was used to crosscheck the total energies reported in this work along with the underlying population bias
associated with the pair-branching algorithm. The reported total energies have negligible population bias and time step error.

\section{Results and Discussion}
For numerical examples, 
we picked a series of one-dimensional (1D) hydrogen chains (H-chains) and water clusters which clearly illustrate the following:
\begin{enumerate}
\item[(1)] The VR technique discussed in \cref{sec:cv} is very effective.
\insertrev{The VR was found to be so effective with restricted Hartree-Fock trial wavefunctions for systems considered here, which allowed
for a very small number of stochastic orbitals to sample (i.e., $N_\xi=1$).}
\item[(2)] The overhead of using CD-sRI with VR for computing ph-AFQMC total energies as opposed to using CD is negligible, unlike for the THC and LR approaches.
\item[(3)] CD-sRI is a cubic-scaling local energy algorithm if implemented correctly \insertrev{(i.e., with a proper tensor contraction ordering)}.
\end{enumerate}
(1) and (2) are rather difficult to show analytically and therefore we provide 
numerical examples to support our claims. While (3) was formally shown in \cref{sec:CD-sRI}, we provide timing benchmarks to bolster support for this analysis.
For 1D H-chains, the inter-hydrogen distance was fixed to be 1.6 Bohr and the water cluster geometries were taken from ref. \citenum{lee2019systematically}.

\begin{figure}[h!]
\includegraphics[scale=0.85]{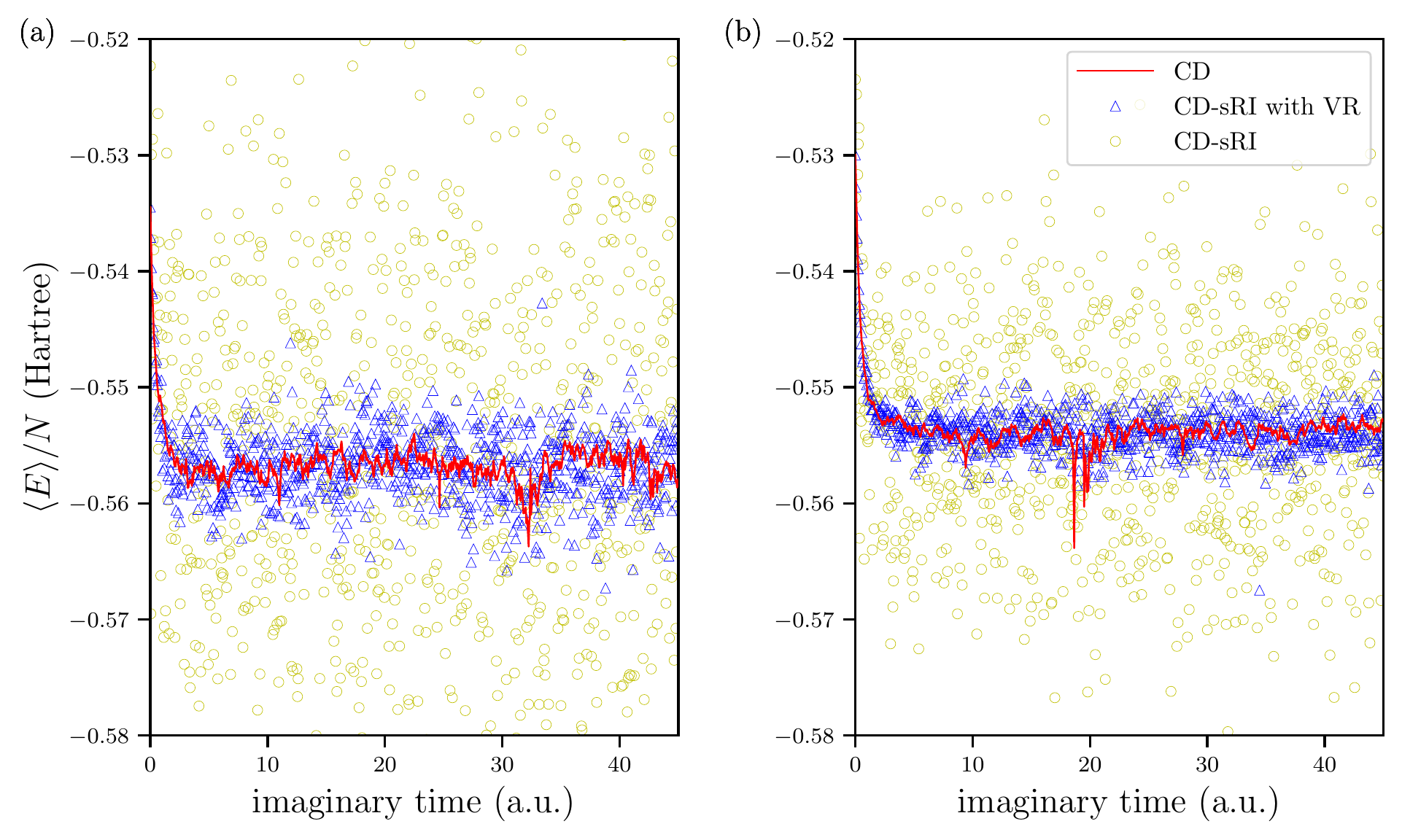}
\caption{\label{fig:hchain}
An example of 
imaginary-time propagation trajectories (energy per H atom as a function of imaginary time) for (a) \ce{H10} and (b) \ce{H40} with the cc-pVDZ basis. The energy estimator was evaluated via three different algorithms.
}
\end{figure}

We focus first on
the effect of
the VR technique discussed in \cref{eq:cdsrifinal}.
In \cref{fig:hchain}, we present 
the energy per H atom as a function of imaginary time. 
In both \ce{H10} and \ce{H40}, we observe the same trend.
Without VR, CD-sRI fluctuates significantly around the CD results
while CD-sRI with VR behaves statistically very similarly to CD.
These results indicate that the overhead of CD-sRI with VR is almost negligible 
and this becomes more evident when looking at
statistically averaged total energies.

\begin{table}[h!]
  \centering
  \begin{tabular}{|c|r|r|r|r|}\hline
  & \multicolumn{2}{c|}{CD} & \multicolumn{2}{c|}{CD-sRI with VR}  \\ \hline
$N$ & \multicolumn{1}{c|}{$\langle E\rangle$} & $N_\text{sample}$ 
        & \multicolumn{1}{c|}{$\langle E\rangle$} & $N_\text{sample}$ \\ \hline 
10 & -5.571(1) & 4000  & -5.5696(9) & 4000 \\ \hline
20 & -11.0990(6) & 21506  & -11.0985(6) & 22191  \\ \hline
40 & -22.1612(7) & 48127 & -22.1610(9) & 27942 \\ \hline
80 & -44.2895(7) & 138429  & -44.2890(8) & 91234  \\ \hline
  \end{tabular}
  \caption{
The total energy ($E_h$) of H-chains for $N=10,20,40,80$ using CD and CD-sRI with VR.
The number of statistical samples $N_\text{sample}$ is also given.
}
  \label{tab:hchain}
\end{table}

In \cref{tab:hchain}, we compare CD and CD-sRI with VR total energies for H-chains.
The difference in total energies between two algorithms is
statistically insignificant since they are all within error bars. 
This is not surprising since CD-sRI is not expected to introduce any biases into the estimator and
should recover the CD result with enough samples.
Furthermore, the magnitude of statistical error is similar when comparing the two algorithms for a similar number of statistical samples.

\begin{figure}[h!]
\includegraphics[scale=0.85]{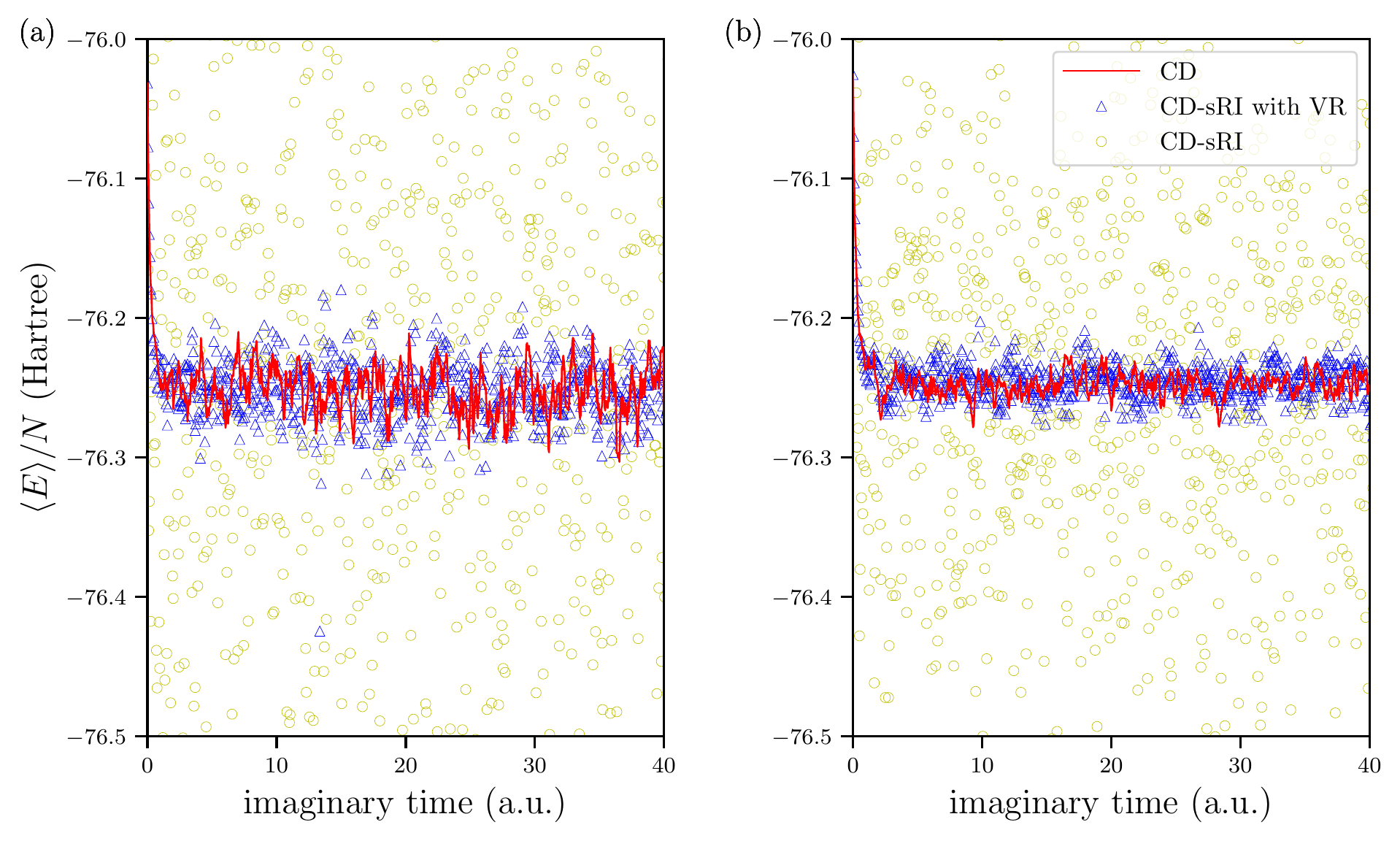}
\caption{\label{fig:h2o}
An example of 
imaginary-time propagation trajectories (energy per \ce{H2O} as a function of imaginary time) for (a) \ce{(H2O)_2} and (b) \ce{(H2O)_8} with the cc-pVDZ basis. The energy estimator was evaluated via three different algorithms.
The number of walkers was 160 in all data.
}
\end{figure}

\begin{table}[h!]
  \centering
  \begin{tabular}{|c|r|r|r|r|}\hline
  & \multicolumn{2}{c|}{CD} & \multicolumn{2}{c|}{CD-sRI with VR}  \\ \hline
$N$ & \multicolumn{1}{c|}{$\langle E\rangle$} & $N_\text{sample}$ 
        & \multicolumn{1}{c|}{$\langle E\rangle$} & $N_\text{sample}$ \\ \hline 
  2 & -152.4987(9) & 16000 & -152.4976(9) & 16000 \\ \hline
  4 & -305.033(1)& 16000 &-305.033(1) &16000 \\ \hline
  8 & -609.989(1) & 49855 & -609.9905(9) & 69596 \\ \hline
  \end{tabular}
  \caption{
The total energy ($E_h$) of water clusters \ce{(H2O)_N} for $N=2,4,8$ using CD and CD-sRI with VR.
The number of statistical samples $N_\text{sample}$ is also given.
}
  \label{tab:h2o}
\end{table}

We have also carried out similar numerical experiments for water clusters, \ce{(H2O)_$N$}, with $N=2,4,8$.
A representative imaginary-time propagation trajectory is given in \cref{fig:h2o}.
The conclusions drawn from the H-chain results holds for these cases as well.
With VR, the statistical fluctuations of CD-sRI are close enough to CD.
This is true for all values of $N=2,4,8$ and we expect this to hold for larger clusters. 
Lastly, we present the absolute energies of \ce{(H2O)_$N$} computed via CD and CD-sRI in \cref{tab:h2o}.
The total energies from two different algorithms lie within their respective error bars.
Similarly to \cref{tab:hchain}, two methods exhibit comparable magnitude of error bars given a similar number of statistical samples.
This strongly suggests that CD-sRI with VR performs as well as CD {\it without overhead}.

\begin{figure}[h!]
\includegraphics[scale=0.85]{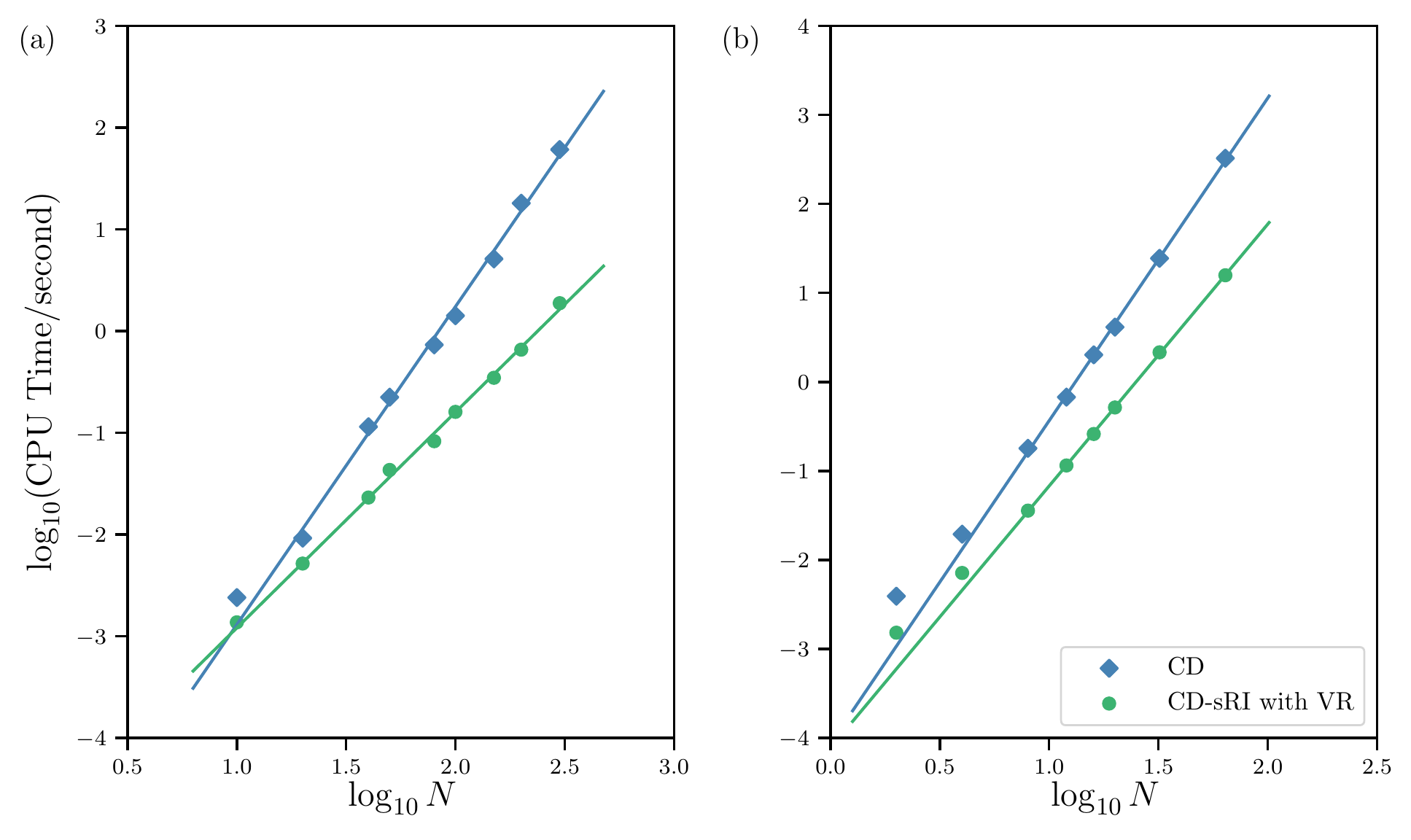}
\caption{\label{fig:timing}
Measured CPU timing for \insertrev{a single local energy evaluation of a single walker} using CD and CD-sRI with VR in the case of 
(a) 1D H-chains ($N=
10,
20,
40,
50,
80,
100,
150,
200,
300
$)
(b) water clusters ($N=2, 4, 8, 12, 16, 20, 32, 64$). 
The slope of CD is 
3.13 for (a) and 3.63 for (b) while
the slope of CD-sRI is 2.12 for (a) and 2.94 for (b).
}
\end{figure}
We finish our discussion with timing benchmarks \insertrev{walker memory saving} of H-chains and water clusters.
For a consistent increase in the number of Cholesky vectors, we employ the density-fitting approximation instead.
The auxiliary basis set used here is that of  Weigend {\em et al.}\cite{Weigend2002}. 
The timing benchmark here was obtained from an Intel(R) Xeon(R) Platinum 8268 CPU 2.90GHz processor and 
only a single thread was used. \insertrev{Furthermore, a single local energy evaluation of a single walker is considered for the purpose of demonstration.}
The timing results are reported in \cref{fig:timing}. Panel (a) shows results for for 1D H-chains up to $N=300$. 
The observed scaling for CD is $\mathcal O(N^{3.13})$ whereas it is $\mathcal O(N^{2.12})$ for CD-sRI (estimated with $R^2$ greater than 0.99).
For the obtained data points, there is no crossover between CD and CD-sRI. CD-sRI is {\it always} faster than CD and
there is no overhead for using CD-sRI compared to CD for a single sample.
Similarly, panel (b) shows the timing benchmark for water clusters up to $N=64$.
The observed scaling for CD is $\mathcal O(N^{3.63})$ whereas it is $\mathcal O(N^{2.94})$ for CD-sRI (estimated with $R^2$ greater than 0.999). The CD algorithm does not exhibit asymptotic scaling for these system sizes, which could be due to the efficient use of BLAS routines.
CD-sRI appears to be quite close to its theoretically predicted asymptotic scaling.
\insertrev{The memory results are obtained based on \cref{tab:memory} and are shown in \cref{fig:memory}.
Even for the smallest systems examined, CD-sRI requires an order of magnitude less memory for each walker.
Due to the scaling differences between CD and CD-sRI, the difference in the memory usage becomes only larger as the system size increases.
This highlights the utility of CD-sRI in reducing the additional memory usage for each walker.
\begin{figure}[h!]
\includegraphics[scale=0.85]{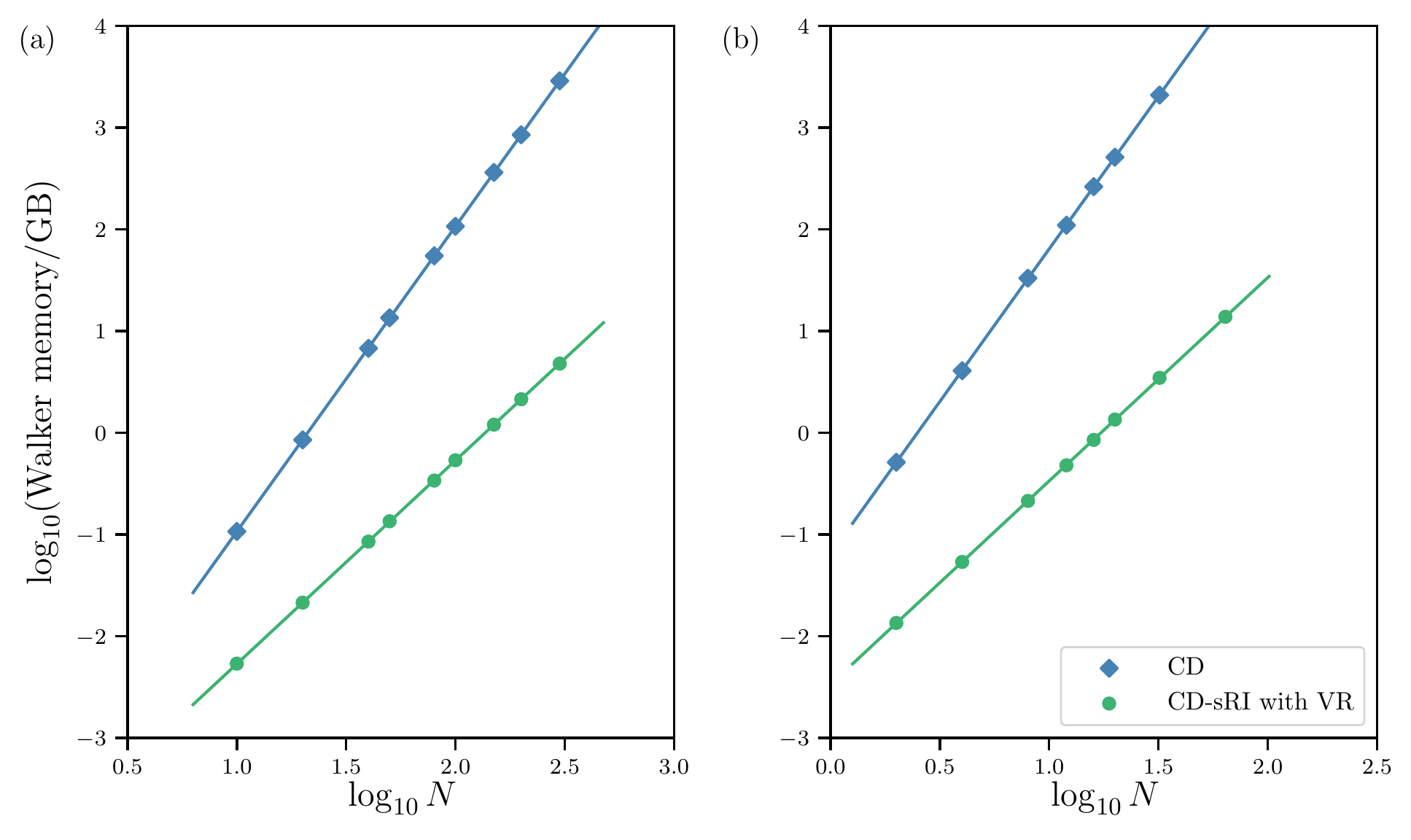}
\caption{\label{fig:memory}
\insertrev{Measured additional walker-specific memory usage for a single local energy evaluation of a single walker using CD and CD-sRI with VR in the case of 
(a) 1D H-chains ($N=
10,
20,
40,
50,
80,
100,
150,
200,
300
$)
(b) water clusters ($N=2, 4, 8, 12, 16, 20, 32, 64$). 
The slope of CD is 
3.0 for both (a) and (b) while
the slope of CD-sRI is 2.0 for both (a) and (b).}
}
\end{figure}
}

In summary, based on simple benchmarks on 1D H-chains and water clusters,
we have successfully demonstrated the
utility of the CD-sRI algorithm.
With the variance reduction technique introduced here,
the CD-sRI algorithm
achieves a cubic-scaling local energy evaluation {\it without overhead}.

\section{Conclusions}
In this work,
we have shown how the stochastic resolution-of-the-identity (sRI) technique
developed by Takeshita {\em et al.}
can be seamlessly integrated with the local energy evaluation of phaseless auxiliary-field Quantum Monte Carlo (ph-AFQMC).
We considered four different existing local energy evaluation strategies: a half-rotated (HR) electron repulsion integral tensor approach,  a Cholesky decomposition-based approach (CD), tensor hypercontraction (THC), and low-rank (LR) factorization approaches.
We have carefully analyzed their formal scalings and discussed possible scaling reduction \joonho{as well as walker-specific memory reduction} when combined with sRI.

We found that HR-sRI neither achieves scaling reduction \joonho{nor walker-specific memory reduction}. It scales quartically with system size like the original HR approach \joonho{with the same memory usage}. On the other hand, the CD approach, which formally scales quartically with system size, can be reduced to cubic-scaling when combined with sRI (CD-sRI). 
\joonho{Furthermore, the walker-specific memory usage is reduced to quadratic from cubic.} Similarly, the cubic-scaling approaches, THC and LR, can be reduced to quadratic-scaling with sRI. 
\joonho{The additional walker-specific memory requirement is also reduced from quadratic to linear.} Without sRI, previously available algorithms using THC and LR achieve cubic scaling, but the overhead associated with them limits the applicability of these algorithms. 
Therefore, we have examined a new cubic-scaling algorithm, CD-sRI, with a particular focus on characterizing the overhead associated with the algorithm.

We applied the CD-sRI algorithm to one-dimensional hydrogen chains (1D H-chains) and water clusters.
With a variance reduction (VR) technique developed for CD-sRI, CD-sRI with VR exhibits no overhead compared to CD 
up to \ce{H80} and up to \ce{(H2O)_8}. By no overhead, we mean that CD-sRI and CD exhibit a similarly large error bar for a similar number of statistical samples. Therefore, CD-sRI adds no additional cost for computing the total ph-AFQMC energy for a fixed statistical error.
Furthermore, we have performed a timing benchmark for H-chains up to $\ce{H300}$ and water clusters up to $\ce{(H2O)_{64}}$. 
We observed reduced scaling from CD-sRI compared to CD. Specifically, the observed scaling of CD was $\mathcal O(N^{3.13})$ for H-chains and $\mathcal O(N^{3.63})$ for water clusters while we observed $\mathcal O(N^{2.12})$ for H-chains and $\mathcal O(N^{2.94})$ for water clusters in the case of CD-sRI. Based on these observations, we argue that the CD-sRI algorithm should be the standard cubic-scaling local energy algorithm with an expected generic speed-up compared to the conventional CD algorithm.

The remaining challenges for large-scale ph-AFQMC simulations are the memory bottleneck and the cubic-scaling walker propagation that occurs every time step. Both THC and LR approaches have successfully addressed the memory bottleneck\cite{malone_isdf,motta_thc} as they require only quadratic-scaling storage for each walker. More theoretical development is needed for accelerating the walker propagation. 
With THC-sRI or LR-sRI, the local energy evaluation scales quadratically with system size.
In these cases the bottleneck for ph-AFQMC is the walker propagation. 
It has been shown that a significant speed-up of the propagation is possible by utilizing graphical processing units (GPUs)\cite{shee2018gpu,malone2020gpu}. \joonho{Furthermore, the significant reduction in the requirements for walker-specific extra memory will be particularly useful with GPUs since in this case the available memory is highly limited.}
\joonho{With sRI, the propagation is the bottleneck in AFQMC for all system sizes. Therefore, it will be interesting to explore ideas like sRI to design new algorithms for walker propagation that scale quadratically (or less) with system size.}
We  reserve this and the exploration of additional projects centered on THC-sRI or LR-sRI with GPUs for large-scale ph-AFQMC applications for future work.

\section{Acknowledgement}
We are indebted to Fionn Malone for stimulating discussions and pointing out the potential usefulness of control variate in reducing the variance.
We also thank Fionn Malone for help with QMCPACK\cite{qmcpack,qmcpack2} which was used to verify
the lack of population bias of reported numbers in this work.
We thank Evgeny Epifanovsky for useful discussions on efficient implementations of the local energy evaluation. 
\joonho{We thank Miguel Morales and Shiwei Zhang for helpful comments.}
D.R.R. acknowledges support from grant NSF-CHE 1954791.

\section{Data Availability}
The data that support the findings of this study are available
from the corresponding author upon reasonable request.

\bibliography{sri_afqmc}
\end{document}